\documentclass{IEEEcsmag}

\usepackage[colorlinks,urlcolor=blue,linkcolor=blue,citecolor=blue]{hyperref}
\expandafter\def\expandafter\UrlBreaks\expandafter{\UrlBreaks\do\/\do\*\do\-\do\~\do\'\do\"\do\-}
\usepackage{upmath,color}

\jvol{XX}
\jnum{XX}
\paper{8}
\jmonth{Month}
\jname{Publication Name}
\jtitle{Publication Title}
\pubyear{2021}

\usepackage[]{xcolor}
\usepackage{tikz}
\newcommand*\circled[1]{\tikz[baseline=(char.base)]{
            \node[shape=circle,draw,inner sep=.5pt] (char) {#1};}}

\definecolor{fancolor}{HTML}{dd3497}

\definecolor{jiayicolor}{HTML}{3182bd}

\definecolor{mascarocolor}{HTML}{f03b20}

\definecolor{rosscolor}{HTML}{3182bd}

\newcommand{\addition}[0]{}

\begin{document}

\sptitle{Article Type: Feature}

\title{FEWSim: A Visual Analytic Framework for Exploring the Nexus of Food-Energy-Water Simulations}

\author{Fan Lei}
\affil{Arizona State University, Tempe, AZ, 85281, USA}

\author{David~A.~Sampson}
\affil{Arizona State University, Tempe, AZ, 85281, USA}

\author{Jiayi Hong}
\affil{Arizona State University, Tempe, AZ, 85281, USA}

\author{Yuxin Ma}
\affil{Southern University of Science and Technology, Shenzhen, 518055, China}

\author{Giuseppe Mascaro}
\affil{Arizona State University, Tempe, AZ, 85281, USA}

\author{Dave White}
\affil{Arizona State University, Tempe, AZ, 85281, USA}

\author{Rimjhim Agarwal}
\affil{Arizona State University, Tempe, AZ, 85281, USA}

\author{Ross Maciejewski}
\affil{Arizona State University, Tempe, AZ, 85281, USA}

\markboth{THEME/FEATURE/DEPARTMENT}{THEME/FEATURE/DEPARTMENT}

\begin{abstract}\looseness-1The interdependencies of food, energy, and water (FEW) systems create a nexus opportunity to explore the strengths and vulnerabilities of individual and cross-sector interactions within FEW systems. However, the variables quantifying nexus interactions are hard to observe, which hinders the cross-sector analysis. To overcome such challenges, we present FEWSim, a visual analytics framework designed to support domain experts in exploring and interpreting simulation results from a coupled FEW model. FEWSim employs a three-layer asynchronous architecture: the model layer integrates food, energy, and water models to simulate the FEW nexus; the middleware layer manages scenario configuration and execution; and the visualization layer provides interactive visual exploration of simulated time-series results across FEW sectors. The visualization layer further facilitates the exploration across multiple scenarios and evaluates scenario differences in performance using sustainability indices of the FEW nexus. We demonstrate the utility of FEWSim through a case study for the Phoenix Active Management Area (AMA) in Arizona.
\end{abstract}

\maketitle

\chapteri{R}esponding to social and environmental challenges such as population growth, climate change, human dietary requirements, etc., requires integrative computer-based approaches to support policy and management decision-making for food, energy, and water (FEW) systems. The complexity of FEW systems, along with the uncertainty of social and environmental drivers that impact food, energy, and water sustainability and resilience, requires a nexus approach \cite{KARAN201886}.
For instance, within such a FEW nexus approach, water plays a crucial role in land irrigation for food and in cooling power plants for energy production. Energy is integral to agriculture procedures and to water treatment. In addition, agricultural activities produce energy through biofuels but affect water quality. Many variables quantifying nexus interactions are difficult to observe \cite{leck2015tracing}; therefore, we need models to analyze and interpret these complex networks. By employing a modeling framework coupled with policy levers, we can untangle these intricate relationships and drive intervention points where success criteria can be monitored using appropriate metrics (e.g., sustainability or resilience indices)~\cite{yadav2021integrated}. Within the framework, systems dynamics models, with functional dependencies, address the study of complex relationships and the inherent feedback of FEW systems, helping to uncover unintended consequences and potential synergies across the FEW sectors \cite{newell201940}.

Simulation systems, such as the Water Evaluation and Planning (WEAP) and the Long-range Energy Alternatives Planning (LEAP), only provide limited coupling features of sharing simulation outputs between the two systems. However, users still need to go back and forth between the two system interfaces when conducting cross-sector analysis, citing the need for a unified exploration environment where all variables and simulation information can be accessed and displayed in a single interface. The literature is sparse regarding studies that have analyzed the simultaneous inter-dependencies among all three sectors of a FEW system. Many existing works focus on characterizing variations in a single sector~\cite{chengguo2023water} or simple interplay between factors in two sectors (e.g. \cite{guan2023impacts, mounir2021investigating, guan2020metropolitan}). The sector-specific processes of FEW systems require the use of multiple coupled models to assess interactions in the linked system. In this case, a coupled framework could be structured to examine specific place-based policies as they influence the three integrative sectors. Fernando Miralles-Wilhelm \cite{motesharrei2016modeling} concedes that previous efforts lacked the methodological components necessary to conduct an integrated policy analysis, and Liu et al.~\cite{liu2017challenges} suggests that integrating tools and software packages would help people improve their understanding of the FEW nexus. A comprehensive interactive visualization component is needed to reveal insights into the FEW nexus simulation results to support decision-making and policy-making for stakeholders. 

In this paper, we propose a visual analytics framework, FEWSim, to support the exploration and analyses of simulation outputs from a coupled food, energy, and water systems dynamics simulation model. It is a comprehensive, interactive visualization framework that stakeholders can use to stimulate questions and provide answers regarding the FEW management and resilience of the study area. To accomplish this, we: \textbf{1)} design a web-based framework with a middleware system that connects to the coupled FEW nexus models and asynchronously links the simulation results to the visualization interface through a database (users can use the middleware interface to create and select interesting scenarios based on different combinations of input variables, e.g., climate files, policy choices, etc.); \textbf{2)} use a parameterized, coupled model to run \addition{real-world} case study simulations for the study area; \textbf{3)} incorporate dynamic visualizations into an interface that reads the simulation outputs from a database, and; \textbf{4)} enable the system to compare simulated variables in specific/multiple scenarios among the FEW sectors. \addition{We demonstrate the design and use of FEWSim through a real-world analytical case in the Phoenix Active Management Area (AMA), showcasing how the framework supports FEW nexus analysis and facilitates stakeholder engagement efforts initiated by White et al.~\cite{white2017stakeholder}.}

\begin{figure}[t]
	\centering
	\includegraphics[width=\linewidth]{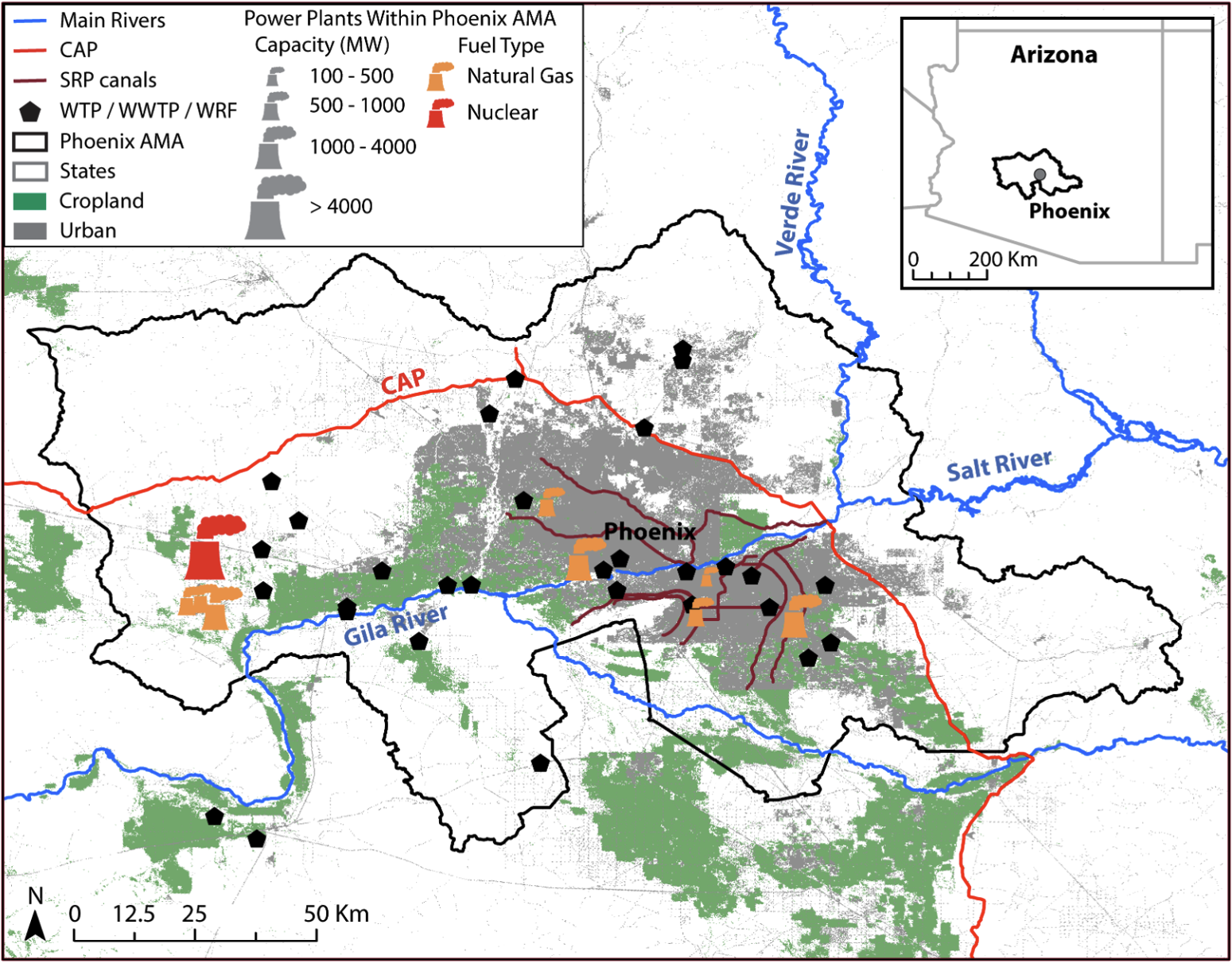}
	\caption{The Phoenix Active Management Area (AMA), along with the location of water and energy demand and supply nodes. The Central Arizona Project (CAP) aqueduct; canals of the Salt River Project (SRP); and cropland and urban areas of the Phoenix metropolitan area are included.}
    \vspace{-4mm}
	\label{fig:ama}
\end{figure}

\section{Background}
\label{sec:background}

\subsection{Study Area Overview}
\vspace{1mm}

Our study area is the Phoenix Active Management Area (AMA), a geopolitical management unit defined by the Arizona Department of Water Resources (ADWR) to oversee and manage groundwater extraction by residents and municipalities (Figure~\ref{fig:ama}). \addition{The Phoenix AMA is well suited for FEW research for the following reasons: 1) the region has seen rapid growth, with a population increasing from 1.86 million in 1985 to 6.05 million in 2020\footnote{Arizona Office of Economic Opportunity: \href{https://oeo.az.gov/population/census/2020}{https://oeo.az.gov/population/census/2020}}; 2) the AMA has a strong and vibrant agricultural sector that is particularly vulnerable to unplanned growth and drought. Agribusiness accounts for over 40 percent of the water use in the Phoenix AMA\footnote{\href{https://new.azwater.gov/ama/ama-data}{https://new.azwater.gov/ama/ama-data}}. The FEW sectors of the AMA are all sensitive to a growing population. Although the cropland area has steadily decreased since the 1980s, mainly as a response to the displacement of urbanization, it has stabilized at 700 $km^2$ \cite{fan2014characterizing}.}

\begin{figure}[t]
    \centering
    \includegraphics[width=\linewidth]{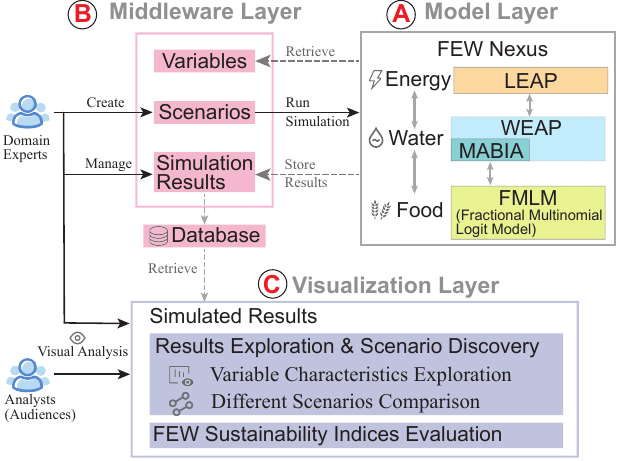}
    \caption{The three-layer asynchronous architecture of the framework. The \textbf{(A) Model Layer} integrates two proprietary models -- WEAP and LEAP, which are employed for water and energy sectors, respectively, along with a statistical model designed to simulate the ``Food'' sector (FMLM). The \textbf{(B) Middleware Layer} exercises control over the model layer by providing scenario creation and management capabilities. 
    A visual analytical interface constitutes the \textbf{(C) Visualization Layer}, which illustrates the simulated outcomes utilizing interactive visual components.}
    \vspace{-4mm}
    \label{fig:architecture}
\end{figure}

\subsection{The FEW Nexus}
\vspace{1mm}


The FEW nexus represents the intricate interconnections between food, energy, and water sectors, highlighting their mutual dependencies and the trade-offs involved in their management. In our framework, we establish the FEW nexus by coupling the food, energy, and water simulation models into one integrated ``\textit{Model Layer}'' as shown in Figure~\ref{fig:architecture}-A.

\vspace{1.5mm} \noindent \textbf{Food Sector: Fractional Multinomial Logit Model (FMLM)}.
We use a fractional multinomial logit model~\cite{Buis2008} to explore scenarios regarding agricultural cropping patterns in the Phoenix AMA, and estimate relative crop land shares (as a proportion between zero and one)~\cite{papke1996econometric}. FMLM estimates for each year the relative area for each crop in the individual irrigation district modeled by WEAP. The predictors employed in this approach rely on variables such as crop yields, prices, as well as climate factors like temperature and precipitation \cite{cho2017climate}. The crops considered in FMLM are: 1) \textit{alfalfa hay}, 2) \textit{barley}, 3) \textit{upland cotton}, 4) \textit{corn}, 5) \textit{spring durum wheat}, and 6) \textit{vegetables}. These crops are chosen because they account for more than 90 percent of the irrigated area in Maricopa County over the study period, and there are sufficient empirical data for model development and assessment. Estimates of crop yield, price received, and acres harvested for the period 1989-2018 were obtained from the USDA's National Agricultural Statistics Service (NASS). In this article, we assume that the county-scale data are representative of the Phoenix AMA because of their average price, yield, and proportional representations. It can also be extended to other areas within the AMA but outside of the county.

\vspace{1.5mm} \noindent \textbf{Energy Sector: LEAP}.
LEAP~\footnote{\href{https://leap.sei.org/default.asp?action=home}{https://leap.sei.org/default.asp?action=home}} is a software package designed to support energy resource planning and management. It simulates electricity generation from diverse fuel types to satisfy demand from different end-users through simple dispatch rules. The model setup includes 1) an energy demand structure based on specific activity levels (e.g., population, water flow) and energy intensities (e.g., per capita or per unit volume of energy consumption) and 2) multiple electricity supply sources with their characteristics, such as fuel type, capacity, percent of energy losses, and reserve margins. Model inputs can be constant or vary in time, while outputs include time series of energy demand from each end-user, and generation and greenhouse gas emissions at each power plant. For our purposes, we use the API for LEAP to have access to model parameters and outputs, and to control simulations using scripts. Mounir et al. \cite{mounir2021investigating} respectively applied LEAP at annual and monthly time scales in the Phoenix AMA to explore future portfolios of energy generation.

The energy supply for our study area includes 9 power plants located within the study area and 27 power plants outside the boundary. These plants produce electricity using a variety of fuels, including coal, natural gas, uranium, and renewable resources (i.e., solar radiation, wind, and water). Transmission and distribution losses were assumed to be 5 percent, while the planned reserve margin was set to 15 percent. The energy demand structure includes residential, commercial, and industrial sectors; within the latter, the water infrastructure sector refers to the electricity demand associated with water transportation and treatment. Mounir et al. describe additional information on model assumptions and structural attributes~\cite{mounir2021investigating}.

\vspace{1.5mm} \noindent \textbf{Water Sector: WEAP and MABIA}.
WEAP~\footnote{\href{https://www.weap21.org/}{https://www.weap21.org/}} is a water management model that simulates water allocations in a network of supply sources and demand nodes under infrastructure and policy constraints. A linear programming heuristic permits scenario-driven simulation to optimize water demand and environmental requirements for water. In our application, we designed a network of four supply sources (CAP canal, Salt River Project (SRP) reservoir system, groundwater, and reclaimed) that deliver water to the municipal, agricultural, industrial, power plants, and Native American demand sectors. Monthly flows from the SRP reservoir systems and CAP were externally provided to WEAP. We also input values that depend on a user-defined scenario of future water availability. For the agricultural sector, we identified 12 irrigation districts and applied the MABIA crop module to simulate irrigation and yield as a function of climate variables. The municipal, industrial and Native American demands were obtained as a function of population and water intensities (e.g., domestic per capita water demand). WEAP:MABIA was calibrated and tested for the Phoenix AMA on monthly~\cite{guan2023impacts} and annual~\cite{guan2020metropolitan} time scales over the periods 1985-2009 and 2008-2018. Model setup and testing were conducted using multiple datasets from the Arizona Department of Water Resources~\footnote{\href{https://azwatermaps.azwater.gov/wellreg/}{https://azwatermaps.azwater.gov/wellreg/}}, US Department of Agriculture, CAP and SRP reports, and Arizona Meteorological Network (AZMET).

\vspace{1.5mm} \noindent \textbf{Model Layer: Coupling Models}.
Our framework contains the functionality of managing and running a coupled WEAP:MABIA-LEAP model with the addition of the FMLM as a ``\textit{Model Layer}''. This layer contains a series of Python scripts that execute each model using inherent APIs, one at a time, with WEAP:MABIA and LEAP communicating with each other to update the system states. Depending on the desired application, this framework permits the use of pre-processed or real-time outputs from the FMLM model. When using real-time outputs, the FMLM was initiated in 1990 (using 30 years of empirical data), and it runs through 2050.
Following either approach, we extract the outputs for 2015 through 2050 and input them into WEAP:MABIA to update the relative crop area for each crop and each irrigation district as a time series. 
We first run WEAP:MABIA to estimate water allocations to all sectors along with crop production in irrigation districts. At each time step, ``links'' are created between the two models to exchange information, as follows: (1) outputs of WEAP are used as input into LEAP to update the electricity demand associated with the water infrastructure sector, and (2) outputs of LEAP are used by WEAP to specify the water demand needed to fulfill the generation of electricity at power plants. The model coupling procedure was developed in close collaboration with domain experts. This approach is consistent with the methodologies described in Guan et al.'s nexus research~\cite{guan2023impacts, guan2020metropolitan}.

\section{Design Overview}
\label{sec:design_overview}

\subsection{Design Goals}


During the design process, we collaborated closely with domain experts specializing in simulation modeling across the food, water, and energy sectors. Through regular meetings \addition{and formative interviews,} we identified key analytical tasks and distilled them into design requirements tailored to support their daily workflows. \addition{To support a clear and focused demonstration of the framework’s design and implementation, we anchored our development in a representative analysis of the Phoenix Active Management Area (AMA). This case, based on expert input, centers on evaluating the impacts of incrementally increasing water use efficiency (WUE), energy use efficiency (EUE), and irrigation efficiency (IE). It reflects typical workflows in which analysts construct scenarios by adjusting key parameters while holding others constant. Using this case allowed us to illustrate the functionality and interaction design of the framework in a real-world context, and to facilitate expert-driven evaluation aligned with domain practices.}

\vspace{1.5mm} \noindent \textbf{Integrate the Models between Sectors.} The traditional simulation methods separate the three sectors into individual models and handle the integration of results via post-processing. However, such methods limit the modeling of the interplay between different sectors which has the potential to produce more comprehensive results by integrating the information from multiple aspects in the same modeling process. Thus, an integrated simulation process should be supported in our framework for analyzing the FEW nexus.

\vspace{1.5mm} \noindent \textbf{Manage, Present, and Explore Multiple Scenarios.} Existing simulation toolkits primarily focus on supporting accurate simulations and modeling. However, integrated exploration and analytics should facilitate scenario management, presentation, and analytics. In such an environment, analysts should be able to create new scenarios, load existing scenarios, and start simulations on demand. Effective presentations and scenario-wise comparisons of simulation results and evaluation measures should also be supported to enhance exploration and knowledge discovery of the simulation results.

\subsection{Analytical Tasks}
The first design goal is supported by previous research \cite{mounir2021investigating} \addition{on} coupling simulation models. \addition{Building upon this, our framework integrates parameterized coupled simulations. The second goal involves developing a visual analytics system tailored for FEW nexus analysis. To effectively achieve these objectives, we defined two new analytical tasks (T1 and T2) based on our comprehensive literature review, direct interviews with domain experts specializing in FEW nexus modeling, and input gathered through stakeholder workshops described by White et al.~\cite{white2017stakeholder}. It is important to note that tasks T1 and T2 are original contributions specifically addressing visualization and analytical needs identified during this research process.} 

\vspace{1.5mm} \noindent \textbf{T1: Scenario Creation and Management.}
Traditional simulation scenario creations often configure the variables for each model independently within their respective software environments and manually integrate the results. This laborious process is also constrained by the high time complexity, making it inherently inflexible. There is a pressing need for a flexible scenario creation interface accompanied by guided and on-demand variable filtering and selection. This includes:
\vspace{-1mm}
\begin{itemize}
    \item \textbf{T1.1} Retrieve and List all the variables in the three sectors in an organized manner;
    \item \textbf{T1.2} Create new scenarios by setting the input variables and running the simulations;
    \item \textbf{T1.3} Store and manage (edit and remove) all simulated scenarios.
\end{itemize}

\vspace{1.5mm} \noindent \textbf{T2: Results Exploration and Scenario Discovery.}
After the simulation results are retrieved from the models, interactive visual exploration and evaluation of the simulation results among the FEW sections would be required. This may include the ability to:
\vspace{-1mm}
\begin{itemize}
    \item \textbf{T2.1} Inspect the characteristics of variables in a specific scenario, including hierarchical structures, composition, inter-sector relationships, and values along with the time steps, trends, and outliers;
    \item \textbf{T2.2} Compare different variables in a specific scenario;
    \item \textbf{T2.3} Compare a specific variable across multiple scenarios;
    \item \textbf{T2.4} Explore scenario differences in performance by evaluating the sustainability indices of the FEW nexus.
\end{itemize}

\subsection{Framework Overview}

\addition{Distilled from the analytical tasks and interviews with domain experts reflecting their practical needs, the FEWSim design integrates coupled simulations across multiple FEW sectors and visualization components to support scenario exploration. As illustrated in Figure~\ref{fig:architecture}, our framework comprises three interconnected layers: (A) a model layer, (B) a middleware layer, and (C) a visualization layer. The model layer connects independent simulation models for food (FMLM), water (WEAP:MABIA), and energy (LEAP). The middleware layer manages the collection, configuration, and storage of inputs and outputs from the model layer (\textbf{T1}), offering a graphical user interface for dynamic scenario generation. Finally, the visualization layer presents structured data from the middleware database through interactive visualizations, facilitating scenario analysis and comparison (\textbf{T2}). This layered design aims to support sustainability analysts, policymakers, and stakeholders by providing robust decision-making tools for integrated FEW policy assessment and governance.}

\section{Scenario Creation and Management}
\label{sec:scenario_management}
\vspace{1mm}

The initial step in our analytical pipeline involves configuring and executing simulation scenarios under varied conditions within the FEW nexus, \addition{followed by retrieving the generated data as inputs for the visual analytics interface. Although our model layer integrates the interactions among individual FEW sector models, conventional workflows typically require analysts to manually configure input variables and independently collect modeling outputs within each sector’s respective software environment. This fragmented and labor-intensive approach significantly limits the flexibility necessary for integrated analysis in contemporary visual analytics systems. To overcome this limitation, we have implemented a more flexible middleware mechanism, enabling analysts to dynamically configure input variables across the coupled models, efficiently manage the retrieval of simulation outputs, and facilitate seamless communication with the visual analytics interface. Additionally, the middleware's asynchronous design addresses the significant computational demands of sequentially running the WEAP, LEAP, and Fractional Multinomial Logit Model (FMLM); initializing and executing a single scenario takes approximately 3.54 hours on an Intel i7-7700 CPU system.}


\begin{figure*}[t]
	\centering
	\includegraphics[width=\linewidth]{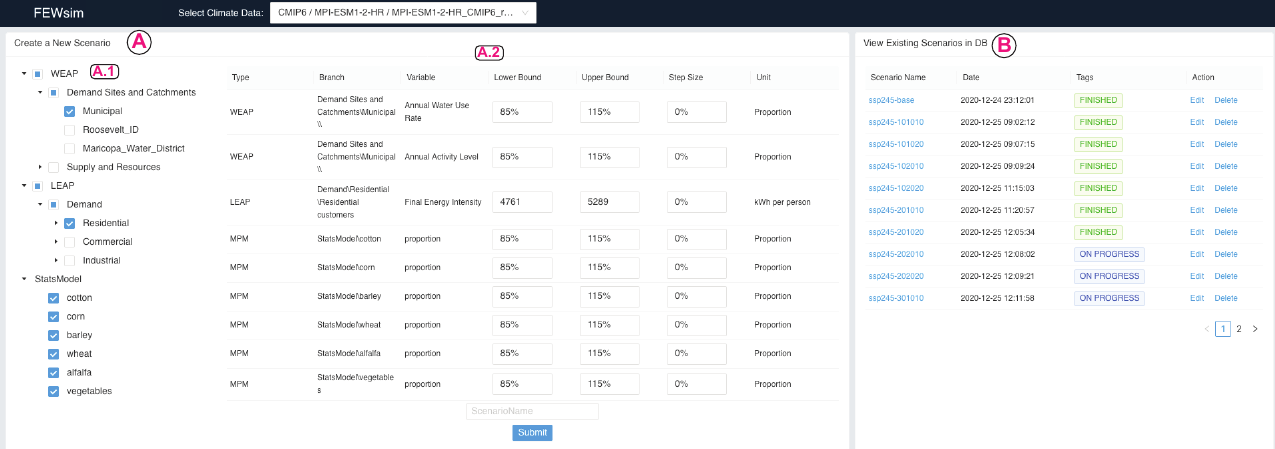}
	\caption{The FEWsim middleware interface. Component (A) permits branch expansion of individual model variables for each FEW sector to enable scenario generation. Analysts can pick important model variables (A.1) and adjust their value, either as single values or as multiple scenario space using minimum and maximum bounds, incremented over a step size that controls the total number of scenarios (A.2). Component (B) presents the simulation scenario status and manages the simulated results.}
    \vspace{-4mm}
	\label{fig:middleware-interface}
\end{figure*}

\subsection{Middleware Layer}

The middleware layer manages and controls the simulation process by communicating with the model layer. This layer also provides encapsulated data storage so that people can see the simulation results through visualizations. There are three main operations supported by the middleware: FEW nexus variables collection (\textbf{T1.1}), scenario creation by editing input variable features (\textbf{T1.2}), and simulation result management (\textbf{T1.3}). We provide an interactive interface with two views to create scenarios by defining the model input variables (Figure~\ref{fig:middleware-interface}-A) and managing the simulated results (Figure~\ref{fig:middleware-interface}-B).

\vspace{1.5mm} \noindent \textbf{Variable Retrieving from the FEW Nexus}.
The middleware initializes the model by detecting and retrieving all variables from WEAP:MABIA and LEAP as shown in Figure \ref{fig:middleware-interface}-A. These variables are arranged into three categories associated respectively with WEAP (water), LEAP (energy), and WEAP:MABIA (agricultural crops and forage and human food for consumption). To facilitate the efficiency and usability of the operations, we arrange the branches under WEAP, LEAP, and FMLM in an interactive tree-structured menu with the same structure as the branches in the WEAP and LEAP software. With the interactive tree menu (Figure~\ref{fig:middleware-interface}-A.1), analysts can filter and select branches, and explore the related variables to create new scenarios. A variable table lists the corresponding variables belonging to each selected category and branch (Figure~\ref{fig:middleware-interface}-A.2). 

\vspace{1.5mm} \noindent \textbf{Scenario Creation}.
After retrieving the branches and the corresponding variables, analysts can create scenarios by adjusting the variables of interest in each model. After selecting target branches, the corresponding variables are displayed in the variable table. In addition to the variable name, type (WEAP/LEAP/ FMLM), unit, and branch path, the table also contains editable variable properties, including the lower bound, upper bound, and step size for a scenario analysis. Analysts can also select the reference or impact the climate scenario. The ``Create scenario'' button starts the simulation process.

To communicate with the model layer, the middleware processes the scenario creation requests into a data format recognizable by WEAP:MABIA and LEAP. We wrap the variable data with related branch information in the JSON format and start the simulation process through the Windows Component Object Model (COM) supported by the WEAP and LEAP software. 
The coupled model first runs the simulation as the base scenario 
without changing any variables. The model then runs the simulation iteratively by increasing or decreasing each user-selected variable value to the selected percentage change as determined by the step size until the value reaches the upper or lower bound. All possible combinations of the user-changed variables in the climate file are simulated.

\vspace{1.5mm} \noindent \textbf{Simulation Result Management}.
When the simulation completes, the middleware extracts the monthly time series results from WEAP and LEAP, and aggregates the results at an annual scale. All results, including the base scenario, are stored in a MongoDB\footnote{https://www.mongodb.com/} database with a user-defined case name. The results are organized into a tree structure based on the hierarchical branches in WEAP:MABIA and LEAP. Analysts can check the status of the simulation (in progress or finished), edit, and delete the simulated scenarios in the result management view (Figure~\ref{fig:middleware-interface}-B). Clicking the ``Edit'' button of a simulated case enables the user to redefine the lower bound, upper bound, and step size of the variables, and re-run the simulation to update the simulation results.

\begin{figure*}[t]
	\centering
	\includegraphics[width=0.95\linewidth]{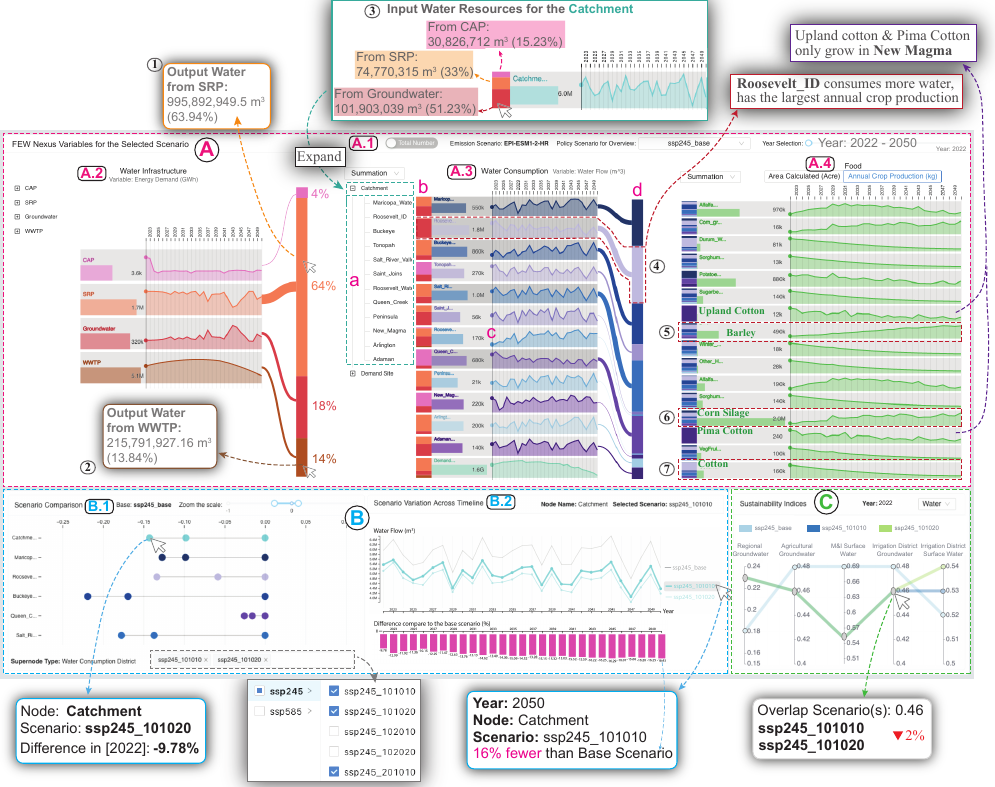}
	\caption{
    The visual analytics interface of FEWSim facilitates the exploration of coupled model simulations for FEW nexus analysis under the base scenario of climate file ssp245. The Coupled Model Variable Exploration module (A) highlights the interconnections among FEW sectors, beginning with the Water Infrastructure sector (A.2, energy consumption), followed by the Water Consumption sector (A.3), and finally the Food sector (A.4), here examined as the agricultural irrigation districts found in our study area. \addition{The Cross-scenario Comparison panel (B) enables analysts to compare percentage changes in node values across selected scenarios for a specific year (B.1) and to examine overall time-series patterns across multiple scenarios within a selected sector (B.2).} Finally, the Sustainability Indices Exploration panel (C) \addition{visualizes proportional differences across selected scenarios for each sustainability index within the FEW nexus.}
    }
    \vspace{-4mm}
	\label{fig:vis-interface}
\end{figure*}

\section{Visual Analytics Interface}
\label{sec:visual_interface}
\vspace{1mm}

Analysts can visually explore the data from the simulated scenarios stored in the middleware through a visual analytics interface noted as the ``\textit{Visualization Layer}'' (Figure~\ref{fig:architecture}-C). The visualization layer consists of three main modules: 1) coupled model variable exploration for a selected scenario (\textbf{T2.1}, \textbf{T2.2}, Figure~\ref{fig:vis-interface}-A), 2) cross-scenario comparison(s) (\textbf{T2.3}, Figure~\ref{fig:vis-interface}-B), and 3) a view of the FEW sustainability index results (\textbf{T2.4}, Figure~\ref{fig:vis-interface}-C). Analysts begin the exploration by selecting a scenario under a certain policy. A year selection slider on the navigation bar controls the year of the time series data in different visualization components.

\subsection{Coupled Model Variable Exploration}

\addition{To support exploration of interdependencies across the FEW nexus, this module provides an interactive visual abstraction of the coupled simulation outputs from LEAP, WEAP, and WEAP:MABIA. It enables analysts to trace how energy demands influence water infrastructure, how water is allocated to different sectors, and how that water supports agricultural production.} To better demonstrate the complex interconnections among the energy, water, and food sectors, inspired by the NEST diagram \cite{8758151}, we combine an alternative design of the Sankey diagram with tree-structure controllers to visually present the relationships between each two sectors and their inner hierarchical structures. 

\begin{figure}[t]
	\centering
	\includegraphics[width=\linewidth]{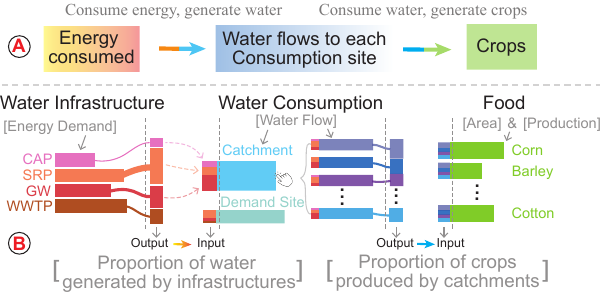}
	\caption{
    The illustration of the FEW nexus linkage. (A) demonstrates the logical relationships between these sectors, while panel (B) presents the structural linkages and key components. These include variable nodes (\textit{Energy Demand}, \textit{Water Flow}, and \textit{Area \& Production} for crops) and output-input inter-sector links within the Coupled Model Variable Exploration module (Figure~\ref{fig:vis-interface}-A).
    }
    \vspace{-4mm}
	\label{fig:few-linkage}
\end{figure}

\vspace{1.5mm}
\noindent \textbf{Energy–Water–Food Linkage.}
\addition{The interface models the FEW system as a network of three interconnected sectors: ``Water Infrastructure,'' ``Water Consumption,'' and ``Food.'' Each sector is treated as a super node, and inter-sector dependencies are represented as directed linkages mediated by shared variables such as energy use, water flow, and crop yield (Figure~\ref{fig:few-linkage}-A). These connections are visually encoded through color-consistent stacked bars and Sankey-style layout conventions.}

\addition{Two key linkages drive the conceptual structure: (1) the energy-to-water linkage shows how different water infrastructures consume energy to produce water, and (2) the water-to-food linkage illustrates how irrigated water supports crop production across districts. Figure~\ref{fig:vis-interface} (A.2–A.4) and Figure~\ref{fig:few-linkage}-B visualize these relationships, with each sector containing coordinated components for interactive exploration.} \addition{This conceptual layout not only clarifies sector-level flows but also supports practical reasoning about sustainability, infrastructure efficiency, and resource dependencies. It is particularly designed to support the exploration of simulated outcomes in scenario configurations relevant to policy.}

\vspace{1.5mm} \noindent \textbf{Branch Characteristics View.}
Each of the FEW sectors has a branch that visually represents the variable characteristics of the FEW model, as well as the hierarchical structure of the branches in each sector (Figure~\ref{fig:vis-interface}-A.2 – A.4). There are four visual components: the \textit{tree menu} (A.3.a), a set of \textit{input resource bars} (A.3.b), a set of \textit{branch value bars} (A.3.c), and an \textit{output stacked bar} that presents the proportional output resource of each sector (A.3.d).

\vspace{1.5mm} \noindent \ieeeguilsinglright
The \underline{tree menu} manages the hierarchical presentation of branches and their subordinate nodes (Figure~\ref{fig:vis-interface}-A.3.a). The branches are arranged according to their original structures in LEAP and WEAP. Analysts can explore the variables for the corresponding branches interactively. The other visual components in the branch characteristics view will be rendered based on the expansion level of the menu.

\vspace{1.5mm} \noindent \ieeeguilsinglright
The \underline{input resource bars} are a set of stacked bar charts that visualize the proportion of input resources for each branch node within a sector for the selected year (Figure~\ref{fig:vis-interface}-A.3.b). \addition{In the ``Water Consumption'' sector, the bars represent how different water infrastructures (e.g., CAP, SRP, WWTP) supply water to catchments; in the ``Food'' sector, they show irrigation contributions by catchment districts. Colors are consistently mapped across sectors, and Sankey-style visual encoding allows analysts to visually trace the flow of resources from origin to destination.} \addition{To support clearer interpretation of small values and input-output consistency, analysts can open a pop-up view for more detailed annotations and enlarged stacked bars.}

\vspace{1.5mm} \noindent \ieeeguilsinglright
The \underline{branch value bars} visualize the values of nodes displayed in the adjacent tree menu (Figure~\ref{fig:vis-interface}-A.3.c). Each bar comprises a horizontal bar and an accompanying line chart. The line chart represents the time-series data for each branch under the selected scenario, with the x-axis indicating the year and the y-axis (line height) denoting the variable's value for each year. The height of the line within each branch row is proportional to its corresponding value. 

By default, the line chart displays the total value for each year for each branch. Analysts can filter the data by selecting specific resources from a drop-down menu if the sector includes input resources. This allows for the exploration of trends and outliers through the line chart. Additional insights are provided via text annotations, which offer explanations about trends and outliers when analysts hover over specific years on the line chart. A circle marker highlights the value of the selected year on the line chart.

The horizontal bar, positioned beside each line chart, represents the value of the selected year, corresponding to the data shown on the line chart. To facilitate easier comparison across branches within a sector, the values of the horizontal bars are normalized using a logarithmic transformation. The x-axis (bar width) indicates the normalized variable values. When analysts apply filters to display a single resource in the time-series line chart, the horizontal bar values automatically update to reflect the selected year's corresponding data. 

Different branch categories within each sector are visually distinguished using distinct colors. Child branches inherit the same color as their parent, maintaining a cohesive visual hierarchy. Audiences can filter the branches based on input resources.

\vspace{1.5mm} \noindent \ieeeguilsinglright
The \underline{output stacked bar} illustrates the proportional distribution of output resources to demand sites within each sector for the selected year (Figure~\ref{fig:vis-interface}-A.3.d). Warm colors represent resources originating from the “Water Infrastructure” sector, while cool colors depict those from irrigation districts within the “Water Consumption” sector. Sankey-style lines connect the stacked bar to the corresponding branch components, visually linking each output to its source (Figure~\ref{fig:few-linkage}-B.Output). \addition{We display numeric labels on hover. These labels explicitly show the proportion of each contributing source, enabling analysts to accurately interpret resource distributions even in cases of low-percentage contributions.}

\vspace{1.5mm}
\noindent \textbf{\addition{Implications for Practice and Design Relevance.}}
\addition{The overall layout enables both high-level overviews and fine-grained exploration. Analysts can use it to compare the role of specific infrastructures across time, detect imbalances in supply-demand flows, and contextualize sustainability metrics within system-level dependencies. The use of color-consistent stacked bars and coordinated views ensures that visual comparisons are intuitive and scalable across sectors and simulation years. The modularity of this design also supports extensibility for additional models or indicators.}

\subsection{Cross-scenario Comparison}

The cross-scenario comparison module \addition{enables analysts to evaluate how the} characteristics of a variable from a selected sector vary across multiple simulation scenarios (\textbf{T2.3}). This module contains two coordinated views: the Scenario Comparison View (Figure~\ref{fig:vis-interface}-B.1), which \addition{supports year-specific comparison of variable values across selected scenarios}, and the Timeline Comparison View (Figure~\ref{fig:vis-interface}-B.2), which \addition{presents the temporal patterns of these variables across scenarios}.


\vspace{1.5mm} \noindent \textbf{Scenario Comparison View.} The scenario comparison view facilitates the comparison of variable values across selected scenarios for branch nodes within the chosen sector, year by year. This view helps analysts examine percentage changes in a variable's value relative to a designated base scenario. Analysts can specify the base scenario via a dropdown menu. Each climate file includes a default base scenario that represents the standard simulation results. To analyze the variations between other simulated scenarios and the base scenario, analysts can select additional scenarios from the hierarchical selection bar within the view.

The framework employs a connected-dot plot to visualize comparisons. We use circles in the same color-encoding that correspond to their branch nodes to represent the selected scenarios. The x-axis indicates the percentage differences in the variable, in the scale of $_-^+100\%$, while each row corresponds to a specific branch node. Gray horizontal lines connect the circles within each row to streamline the visual mapping and enhance comparative analysis. Analysts can adjust the comparison scale by sliding the ``zoom-the-scale'' bar, enabling focused exploration of the visual content. A pop-up panel containing detailed text annotations for each scenario appears when hovering over a circle. Additionally, analysts can explore value variations over time relative to the base scenario by clicking on a scenario circle.  \addition{When multiple variables or inputs are involved in a simulation, this view does not aggregate them. Instead, each scenario represents a unique configuration of simulation inputs (e.g., a particular combination of water use efficiency (WUE) and energy use efficiency(EUE) values), and only one output variable (e.g., Amount of Water Consumption in year 2022) is visualized at a time for accurate, interpretable comparisons.}

\vspace{1.5mm} \noindent \textbf{Timeline Comparison View.} \addition{The view allows analysts to examine temporal trends of the selected variable across multiple scenarios simultaneously within the same interface. This view supports multi-line time series charts, where each line represents one selected scenario, enabling side-by-side inspection of long-term behavior. Clicking a line highlights that scenario and reveals an associated bar chart titled “Difference Compared to the Base Scenario (\%)”. This bar chart displays year-by-year percentage differences between the selected scenario and the base case, offering a precise view of how the scenario deviates over time.}

\addition{The module is intentionally designed to support dynamic scenario selection. Analysts can add or remove scenarios based on their investigative goals, such as comparing the isolated effects of varying WUE while keeping EUE and IE constant. This interactive filtering mechanism ensures focus and clarity during analysis, helping experts explore meaningful differences without visual clutter.}


\subsection{Sustainability Indices Exploration}

The sustainability indices exploration module provides a visual interface for evaluating the sustainability performance of the food, energy, and water sectors, calculated from the simulated variables (\textbf{T2.4}). \addition{A parallel coordinates plot (Figure~\ref{fig:vis-interface}-C) is used to represent proportional differences among selected scenarios across multiple sustainability indicators. Each axis corresponds to a distinct indicator, and each polyline represents one selected scenario, enabling direct visual comparison across the full index space.} \addition{The interactive features of this view allow analysts to highlight specific scenarios, filter using the legend, and explore patterns across indicators through brushing and hovering. This design supports comprehensive multi-scenario comparison, making it easier to detect subtle differences and broader trade-offs between scenarios across the FEW nexus.} \addition{This module enables targeted evaluation of sustainability outcomes based on analysts’ selected scenarios. For example, analysts can explore how increases in water use efficiency (WUE) influence various dimensions of groundwater reliance, or assess how specific policy scenarios impact the balance of energy sourcing and agricultural resource demands. The structure of the visualization supports both holistic and focused sustainability analysis, aligning with diverse expert needs.}

\vspace{1.5mm} \noindent \textbf{Sustainability Index Type.}
The framework defines a set of sustainability indicators across the three FEW sectors, based on expert consultations and established assessment practices~\cite{Foley2011, Sandoval2011}. The water sector includes five indicators: (1) regional groundwater reliance, (2) agricultural groundwater reliance, (3) municipal and industrial (M\&I) surface water reliance, (4) irrigation district groundwater reliance, and (5) irrigation district surface water reliance. The energy sector includes (1) renewable energy use and (2) dependence on electricity imports. The food sector includes (1) agricultural water impact, (2) agricultural energy use, and (3) agricultural carbon emissions.

\subsection{Framework Implementation}

The FEWSim framework is implemented using D3.js and React for the visual components. \addition{To accommodate users with color vision deficiencies, we utilize color schemes selected from colorblind-friendly palettes from ColorBrewer\footnote{\href{https://colorbrewer2.org}{https://colorbrewer2.org}}, supplemented by additional visual encodings such as patterns and interactive highlighting for critical comparison tasks. } The back-end services, such as FEW model coupling and simulation processing and management, are supported by Python and Flask. We use Component Object Model (COM) to handle communications between the coupled model layer and the middleware layer, and the visualization interface retrieves the simulated results from the database through RESTful API. We provide the simulation datasets and the codes of the framework at \href{https://osf.io/p4ezy/}{\texttt{https://osf.io/p4ezy/}}.

\begin{figure}[t]
	\centering
	\includegraphics[width=\linewidth]{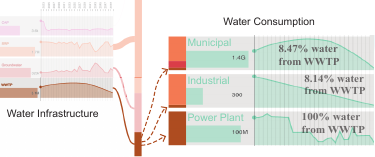}
	\caption{
    Visualization of the water generated by the wastewater treatment plant (WWTP) and its distribution among various water consumption sectors.
    }
    \vspace{-4mm}
	\label{fig:wwtp-case}
\end{figure}
\section{Case Study}
\label{sec:case_study}
\vspace{1mm}

\begin{figure*}[t]
	\centering
	\includegraphics[width=\linewidth]{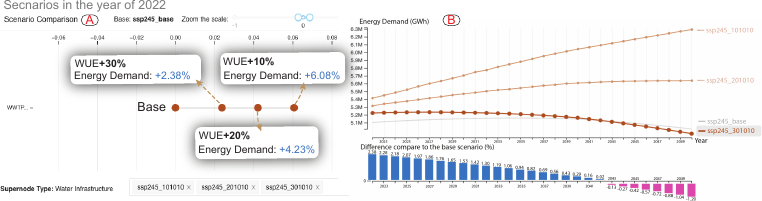}
	\caption{Cross-scenario comparison for the WWTP energy demand between the base scenario and scenarios \addition{where water use efficiency has been increased by 10\% (ssp245\_101010), 20\% (ssp245\_201010), and 30\% (ssp245\_301010).}}
	\label{fig:cross-scenario-comparison}
\end{figure*}

\addition{We present a case study demonstrating how the FEWSim framework facilitates exploratory analysis of the FEW nexus, particularly supporting stakeholder engagement efforts in the Phoenix Active Management Area (AMA), as initiated by White et al.~\cite{white2017stakeholder}.}

\addition{A team comprising domain experts and local stakeholders utilized FEWSim to analyze outputs from a coupled FEW simulation model for the AMA, covering the period 2022–2050. The primary analytic objective was to explore sensitivities of critical variables, municipal water use efficiency (WUE), household energy use efficiency (EUE), and irrigation efficiency (IE), and their interactions within the nexus. Using FEWSim’s scenario management function in the middleware interface, analysts configured new scenarios by incrementally adjusting baseline values of WUE by 10\%, 20\%, and 30\%, and EUE and IE each by 10\% and 20\%. These adjustments were selected through guided filtering and input sliders within the middleware interface, which allowed experts to construct and launch model runs for various parameter combinations (e.g., a “10-10-10” configuration with a 10\% increase in each variable). Analysts then explored the resulting scenario outputs interactively across both baseline and impact climate conditions. For climate context, two scenarios from the MPI-ESM1-2-HR general circulation model under SSP2-4.5 (baseline) and SSP5-8.5 (impact scenario) were used~\cite{guan2023impacts}, representing moderate and intensive emissions scenarios~\cite{o2016shared}.}

\subsection{Energy-water-food Linkage Exploration}

\addition{Experts began by exploring the interconnected dynamics of energy, water, and food sectors using the Coupled Model Variable Exploration module (\textbf{T2.1}). They identified key insights interactively: approximately 63\% of total annual water originates from the SRP system (Figure~\ref{fig:vis-interface}-\circled{1}). Using the Energy-Water-Food linkage visualization, experts quickly discerned that wastewater treatment plants (WWTPs) consume substantial energy relative to their water output (Figure~\ref{fig:vis-interface}-\circled{2}). Further exploration revealed reclaimed water from WWTPs primarily serves municipal, industrial, and power plant sectors, with power plants exclusively sourcing from WWTPs for nearly three decades (Figure~\ref{fig:wwtp-case}). Experts highlighted that irrigation districts predominantly relied on groundwater (51.23\%), prompting deeper exploration into district-specific patterns (Figure~\ref{fig:vis-interface}-\circled{3}). Interactive drill-down into irrigation data from the catchments revealed Roosevelt Irrigation District had the highest water consumption and crop productivity (Figure~\ref{fig:vis-interface}-\circled{4}), while Upland and Pima Cotton were exclusive to the New Magma district.}

\subsection{Variable Characteristics Exploration}

\addition{Leveraging FEWSim’s interactivity, analysts closely examined temporal trends and patterns within sectors (\textbf{T2.2}). Investigations into the Food sector revealed cotton, despite being extensively irrigated, experienced declining productivity from 2022 to 2050, while barley and corn silage production showed consistent increases (Figures~\ref{fig:vis-interface}-\circled{5}, \circled{6}, and \circled{7}). Furthermore, interactive scenario-switching enabled analysts to uncover significant impacts of WUE adjustments on WWTP energy demands and municipal water consumption, highlighting nuanced sectoral interactions.}

\subsection{Cross-scenario Comparison}

\addition{Building upon insights from variable explorations, analysts conducted explicit cross-scenario comparisons to evaluate quantitative impacts of increased WUE on sectoral demands (\textbf{T2.3}). Interactive selection and visualization allowed experts to observe nuanced differences: under a 10\% WUE increase, WWTP energy demands rose significantly over time, whereas a 30\% increase eventually led to reduced energy demands compared to baseline scenarios (Figures~\ref{fig:cross-scenario-comparison}). Analysts also identified notable reductions in irrigation water flows with enhanced irrigation efficiencies, directly quantifying scenario-driven outcomes.}

\subsection{Sustainability Indices Exploration}

\addition{Finally, analysts employed the redesigned Sustainability Indices Exploration view to dynamically compare sustainability metrics across multiple scenarios interactively (\textbf{T2.4}). Utilizing parallel coordinates plots, they efficiently identified key sustainability outcomes: increasing WUE consistently reduced agricultural and irrigation groundwater reliance while variably affecting municipal and industrial surface water reliance. This exploration provided immediate visual evidence of trade-offs and synergies across the nexus. Analysts also confirmed climate scenarios (ssp245 and ssp585) minimally influenced sustainability indices changes, highlighting resilience insights that emerged directly from interactive visual analyses.}

\addition{Overall, this real-world case study explicitly demonstrates the iterative analytic discourse and integrated visual-interactive-computational analyses central to FEWSim, thereby showcasing its unique capabilities in supporting exploratory, stakeholder-driven analyses.}

\begin{figure}[t]
	\centering
	\includegraphics[width=\linewidth]{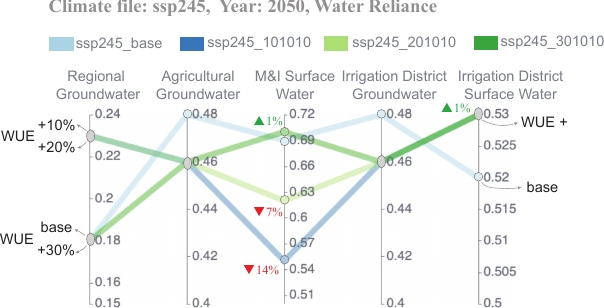}
	\caption{Water sustainability indices comparisons between the base scenario and scenarios that have \addition{increased water use efficiency by 10\% (ssp245\_101010), 20\% (ssp245\_201010), and 30\% (ssp245\_301010) by the year of 2050.}}
    \vspace{-4mm}
	\label{fig:sustainability-comparison}
\end{figure}




\section{Expert Interviews}
\label{sec:evaluation}
\vspace{1mm}

\addition{To evaluate the usability and utility of the FEWSim framework, we conducted semi-structured interviews with three domain experts not involved as co-authors (\textbf{E1–E3}). Informed consent was obtained from all participants prior to their involvement in the interviews. The panel included two specialists in water and energy simulation modeling (\textbf{E1, E2}) and one expert in sustainability assessment and policy analysis within the FEW nexus (\textbf{E3}). All three experts are familiar with the Phoenix Metropolitan Area (AMA), the study area used in our case study, which allowed them to provide contextually grounded feedback on the framework’s design and analysis tasks.}

\addition{Each expert was given approximately 45 minutes to explore FEWSim, engage with the visualization components, and review scenarios demonstrated in the AMA case study. This interactive session was followed by semi-structured interviews guided by key evaluation criteria aligned with the analytical tasks. The interviews were conducted via Zoom, during which experts were encouraged to think aloud and provide immediate feedback while exploring the framework. Following the exploration session, we asked them to share their overall impressions and suggestions. Some experts also submitted additional written feedback afterward.}

\vspace{1.5mm} \noindent \textbf{\addition{Usability of Interactive Visual Components.}}
\addition{Experts provided positive feedback regarding the usability and clarity of the visualization components within FEWSim. \textbf{E1} commended the Timeline Comparison View for enabling efficient cross-scenario comparisons: ``Directly comparing multiple scenarios in the time-series view significantly streamlines analysis by clearly highlighting temporal trends and differences without the cognitive overhead of repeatedly switching scenarios.'' Similarly, \textbf{E3} acknowledged the effectiveness of the parallel coordinates plot for sustainability indices, remarking, ``Adopting parallel coordinates enhances interpretability, as it succinctly visualizes proportional differences across multiple sustainability indicators and clearly distinguishes scenario outcomes.''}

\addition{Additionally, \textbf{E3} specifically appreciated the FEW nexus linkage visualization, noting its balanced presentation of sectoral interactions and detailed variable-level information: ``The visualization thoughtfully integrates high-level inter-sectoral relationships with granular insights into individual variables. This comprehensive representation effectively supports nuanced analyses and decision-making.''}

\addition{However, \textbf{E2} expressed a preference for more explicit visual connections between sectors in the abstracted Sankey design, suggesting, ``Currently, the connections rely heavily on color-coding for mapping flows between sectors. It would be beneficial to explicitly draw visible lines or flow paths to better illustrate sector-to-sector transfers, improving immediate interpretability, especially for analysts less familiar with the color mappings.'' This feedback points towards enhancing clarity in future visualizations.}

\vspace{1.5mm} \noindent \textbf{\addition{Utility in Real-world Analytical Tasks.}}
\addition{The feedback indicated that FEWSim effectively supports real-world analytical tasks. \textbf{E3} affirmed its practical value in decision-making scenarios, commenting, ``The framework matches our typical workflow closely, and the middleware’s asynchronous feature greatly enhances our ability to analyze results incrementally rather than waiting for all simulations to complete.'' \textbf{E2} similarly appreciated FEWSim’s alignment with existing domain practices: ``We usually analyze scenarios incrementally and selectively rather than all at once. This system supports exactly that mode of operation.''}

\addition{However, experts identified potential areas for further enhancement. \textbf{E1} suggested incorporating predictive analytics or machine-learning techniques, stating, ``It would be helpful if the framework could suggest potentially interesting scenarios proactively, based on previously analyzed scenarios.'' \textbf{E3} echoed this sentiment, indicating potential interest in advanced analytical capabilities to support exploratory analysis.}

\addition{The expert feedback thus highlights the framework’s strengths in usability and practical utility, while also pinpointing clear pathways for future improvements, detailed further in the Conclusion and Future Work section.}

\section{Conclusion and Future Work}
\label{sec:conclusion}
\vspace{1mm}

In this work, we present FEWSim, a visual analytics framework that permits scenario creation and visual analysis of the simulation outputs from a coupled food, energy, and water model. The framework consists of a three-layer architecture with rich interactive views that manage the visual analysis for the simulated scenarios asynchronously. The model layer couples the models from the FEW sectors. 
As a communicator between the coupled model layer and the visual analytical interface, the middleware configures the scenario creation, stores, and manages the simulated scenarios. Analysts can create scenarios by selecting a climate file and, if needed, they may also change important input variables either as fixed constants or as changing values across a spectrum of possible values to explore the impact of uncertainty in the inputs on model outputs. 

As the core component, the visual analytical interface retrieves the simulated results from middleware and supports analysts to explore the scenarios from the coupled model simulations. With the novel visual design for the FEW nexus linkages inspired by the abstracted Sankey diagram, analysts can inspect the simulated time-series variables across the FEW sectors. FEWSim also facilitates the exploration across multiple scenarios and evaluates scenario differences in performance using sustainability indices of the FEW nexus. We demonstrate our framework through a practical case study to highlight its potential in supporting FEW nexus analyses.



\subsection{Future Work}

\vspace{1.5mm}
\noindent \textbf{\addition{Scalability and Visual Design Enhancements.}}
\addition{To support analysis across a larger number of simulation scenarios, we plan to enhance the scalability of FEWSim’s visual interface. This includes integrating advanced filtering, grouping, and aggregation techniques that allow analysts to efficiently navigate and compare hundreds or thousands of scenarios. Improving the clarity and responsiveness of comparative views, such as scenario selection panels and time-series overlays, will be a key priority to ensure usability at scale.}

\vspace{1.5mm}
\noindent \textbf{\addition{Machine Learning–Assisted Scenario Recommendation.}}
\addition{To reduce cognitive burden and streamline exploration, we plan to incorporate machine learning–based methods that assist with scenario pre-filtering and recommendation. These algorithms will leverage scenario metadata, analyst-selected parameters, and usage patterns to highlight relevant or anomalous scenarios, enabling more focused and goal-oriented analyses.}

\vspace{1.5mm}
\noindent \textbf{\addition{Quantitative User Evaluation.}}
\addition{While this study reports qualitative insights from expert interviews, we plan to extend the evaluation of FEWSim through quantitative user studies. These evaluations will assess usability, insight generation, and task performance across different analyst profiles using controlled experiments. Results will help validate the system’s design choices and inform future interface refinements.}

\vspace{1.5mm}
\noindent \textbf{\addition{Model Expansion and Standardization.}}
\addition{To improve adaptability across diverse application domains, we plan to integrate additional sector-specific and regional simulation models. This will allow for more granular analysis of FEW dynamics in varied contexts. We also aim to ensure compliance with emerging data standards in the FEW modeling community, improving interoperability and enabling smoother integration with external data systems and tools.}


\def\refname{REFERENCES}

\vspace*{-8pt}

\vspace*{8pt}
\begin{IEEEbiography}{Fan Lei}{\,}
is a PH.D. candidate at Arizona State University. His research interests include data visualization, visual analytics, and information visualization. He received his master's degree in computer science from Durham University. Contact him at flei5@asu.edu.
\end{IEEEbiography}

\begin{IEEEbiography}{David A. Sampson}{\,} is the lead WaterSim model developer for the Decision Center for a Desert City (DCDC). His research interests focus on modeling the municipal water supply and demand. He received a Ph.D. in Forestry: Systems Modeling from Colorado State University. Contact him at darthur.sampson@gmail.com.
\end{IEEEbiography}

\begin{IEEEbiography}{Jiayi Hong}{\,} is Postdoctoral Research Scholar at Arizona State University. Her research interests lie in information visualization. She received her Ph.D. in Computer Science at AVIZ INRIA, Université Paris Saclay. Contact her at jhong76@asu.edu.
\end{IEEEbiography}

\begin{IEEEbiography}{Yuxin Ma}{\,} is an Associate Professor at Southern University of Science and Technology. His research interests are in the areas of visualization and visual analytics. He worked as a Postdoctoral Research Associate at Arizona State University. He received his Ph.D. from Zhejiang University. Contact him at mayx@sustech.edu.cn.
\end{IEEEbiography}

\begin{IEEEbiography}{Giuseppe Mascaro}{\,} is an Associate Professor at Arizona State University. His research interests include stochastic hydrology, watershed modeling, hydroclimatology, climate change, infrastructure modeling, and the food-water-energy nexus. He received his Ph.D. in Hydrology from University of Cagliari. Contact him at gmascaro@asu.edu.
\end{IEEEbiography}

\begin{IEEEbiography}{Dave White}{\,} is a Professor at Arizona State University. His research interests include environmental policy, sustainability, and food, water, and energy nexus. He received his Ph.D. in Forestry from Virginia Tech. Contact him at Dave.White@asu.edu.
\end{IEEEbiography}

\begin{IEEEbiography}{Rimjhim Agarwal}{\,} is a Professor at Arizona State University. Her research interests lie at the interface between sustainability science and international development. She received her Ph.D. in Economics from Cornell University. Contact her at Rimjhim.Aggarwal@asu.edu.
\end{IEEEbiography}

\begin{IEEEbiography}{Ross Maciejewski} {\,} is a Professor at Arizona State University. His primary research interests are in the areas of visualization and explainable AI. He received his Ph.D. in Computer Engineering from Purdue University. Contact him at rmacieje@asu.edu.
\end{IEEEbiography}


\begin{thebibliography}{1}




\bibitem{KARAN201886}Karan, E. \& Asadi, S. (2018). Quantitative modeling of interconnections associated with sustainable food, energy and water (FEW) systems. {\it Journal Of Cleaner Production}, 200, 86-99.




\bibitem{leck2015tracing}Leck, H., Conway, D., Bradshaw, M. \& Rees, J. (2015). Tracing the water–energy–food nexus: Description, theory and practice. {\it Geography Compass}, 9, 445-460.

\bibitem{yadav2021integrated}Yadav, K., Geli, H., Cibils, A., Hayes, M., Fernald, A., Peach, J., Sawalhah, M., Tidwell, V., Johnson, L., Zaied, A. \& Others An integrated food, energy, and water nexus, human well-being, and resilience (FEW-WISE) framework: New Mexico. {\it Frontiers In Environmental Science}. 9 pp. 667018 (2021)

\bibitem{newell201940}Newell, J., Goldstein, B. \& Foster, A. (2019). A 40-year review of food–energy–water nexus literature and its application to the urban scale. {\it Environmental Research Letters}, 14, 073003.







\bibitem{chengguo2023water}Wu, C., Tong, F., Jin, J., Zhou, Y., Nie, B., Cui, Y., \& Zhang, L. (2024). Variation characteristic analysis of regional agricultural water consumption under Budyko-type framework. {\it Hydrological Sciences Journal}, 69(14), 1973–1984. 

\bibitem{guan2023impacts}Guan, X. \& Mascaro, G. (2023). Impacts of climate change on the food-water nexus in central Arizona. {\it Agricultural And Forest Meteorology}, 333, 109413.

\bibitem{mounir2021investigating}Mounir, A., Guan, X. \& Mascaro, G. (2021). Investigating the value of spatiotemporal resolutions and feedback loops in water-energy nexus modeling. {\it Environmental Modelling \& Software}, 145, 105197.


\bibitem{guan2020metropolitan}Guan, X., Mascaro, G., Sampson, D. \& Maciejewski, R. (2020). A metropolitan scale water management analysis of the food-energy-water nexus. {\it Science Of The Total Environment}, 701, 134478.

\bibitem{motesharrei2016modeling}Motesharrei, S., Rivas, J., Kalnay, E., Asrar, G. R., Busalacchi, A. J., Cahalan, R. F., ... \& Zeng, N. (2016). Modeling sustainability: population, inequality, consumption, and bidirectional coupling of the Earth and Human Systems. {\it National Science Review}, 3(4), 470-494.

\bibitem{liu2017challenges}Liu, J., Yang, H., Cudennec, C., Gain, A., Hoff, H., Lawford, R., Qi, J., Strasser, L., Yillia, P. \& Zheng, C. (2017). Challenges in operationalizing the water–energy–food nexus. {\it Hydrological Sciences Journal}, 62, 1714-1720.

\bibitem{white2017stakeholder}White, D., Jones, J., Maciejewski, R., Aggarwal, R. \& Mascaro, G. (2017). Stakeholder analysis for the food-energy-water nexus in Phoenix, Arizona: Implications for nexus governance. {\it Sustainability}, 9, 2204.



\bibitem{fan2014characterizing}Fan, C., Zheng, B., Myint, S. \& Aggarwal, R. (2014). Characterizing changes in cropping patterns using sequential Landsat imagery: An adaptive threshold approach and application to Phoenix, Arizona. {\it International Journal Of Remote Sensing}, 35, 7263-7278.

\bibitem{Buis2008}Buis, Maarten L. "FMLOGIT: Stata module fitting a fractional multinomial logit model by quasi maximum likelihood." {\it Boston College Working Papers in Economics} (2008).

\bibitem{papke1996econometric}Papke, L. \& Wooldridge, J. (1996). Econometric methods for fractional response variables with an application to 401 (k) plan participation rates. {\it Journal Of Applied Econometrics}, 11, 619-632.

\bibitem{cho2017climate}Cho, S. \& McCarl, B. (2017). Climate change influences on crop mix shifts in the United States. {\it Scientific Reports}, 7, 40845.





\bibitem{8758151}Mathis, B., Ma, Y., Mancenido, M. \& Maciejewski, R. (2021). Exploring the Design Space of Sankey Diagrams for the Food-Energy-Water Nexus. {\it IEEE Computer Graphics And Applications}, 41, 25-34.

\bibitem{Sandoval2011}Sandoval-Solis, S., McKinney, D. C., \& Loucks, D. P. (2011). Sustainability index for water resources planning and management. {\it Journal of water resources planning and management}, 137(5), 381-390.

\bibitem{Foley2011}Foley, J. A., Ramankutty, N., Brauman, K. A., Cassidy, E. S., Gerber, J. S., Johnston, M., ... \& Zaks, D. P. (2011). Solutions for a cultivated planet. {Nature}, 478(7369), 337-342.

\bibitem{o2016shared}O'Neill, B. (2016). The Shared Socioeconomic Pathways (SSPs) and their extension and use in impact, adaptation and vulnerability studies.

\bibitem{usta2022assessment}Usta, D., Teymouri, M. \& Chatterjee, U. (2022). Assessment of temperature changes over Iran during the twenty-first century using CMIP6 models under SSP1-26, SSP2-4.5, and SSP5-8.5 scenarios. {\it Arabian Journal Of Geosciences}, 15, 416.







\end{thebibliography}
\end{document}